\documentclass[aps,prb,a4,superscriptaddress,preprintnumbers,showpacs,twoside,twocolumn,floatfix]{revtex4}

\usepackage{graphicx,graphics,color,type1cm}
\usepackage{dcolumn}
\usepackage{amsmath}
\usepackage{amsfonts}
\usepackage{amssymb}
\usepackage{amsthm}
\usepackage{bbm}
\usepackage{setspace}
\usepackage{helvet}
\usepackage{times}
\usepackage{color}

\usepackage{xspace}
\usepackage{epsfig}
\usepackage{graphicx,color,type1cm}
\usepackage{pict2e}
\usepackage{amssymb,wasysym}
\usepackage{nicefrac}
\usepackage{cancel}
\usepackage{bm}

\newlength{\textlarg}

\newcommand{\Sum}{\displaystyle\sum}

\newcommand{\ii}{\mathrm{i}}
\newcommand{\Tr}{\text{Tr}\,}
\newcommand{\diag}{\text{diag}}

\newcommand{\kb}{\textbf{k}}
\DeclareMathOperator{\e}{e}

\begin{document}

\title{Spin-wave analysis of the transverse-field Ising model on the checkerboard lattice}

\author{Louis-Paul Henry}
\affiliation{Laboratoire de Physique, CNRS UMR 5672, Ecole Normale Sup\'erieure de Lyon, Universit\'e de Lyon, 46 All\'ee d'Italie,
Lyon, F-69364, France}
\author{Peter C. W. Holdsworth}
\affiliation{Laboratoire de Physique, CNRS UMR 5672, Ecole Normale Sup\'erieure de Lyon, Universit\'e de Lyon, 46 All\'ee d'Italie,
Lyon, F-69364, France}
\author{Fr\'ed\'eric Mila}
\affiliation{Institut de Th\'eorie des Phenom\`enes Physique, Ecole Polytechnique F\'ed\'erale de Lausanne (EPFL),
CH-1015 Lausanne, Switzerland}
\author{Tommaso Roscilde}
\affiliation{Laboratoire de Physique, CNRS UMR 5672, Ecole Normale Sup\'erieure de Lyon, Universit\'e de Lyon, 46 All\'ee d'Italie,
Lyon, F-69364, France}

\date{\today}

\begin{abstract}
The ground state properties of the $S=1/2$ transverse-field Ising model on the checkerboard lattice
are studied using linear spin wave theory. We consider the general case of different couplings
between nearest neighbors ($J_1$) and next-to-nearest neighbors ($J_2$). In zero field the
system displays a large degeneracy of the ground state, which is exponential in the system
size (for $J_1=J_2$) or in the system's linear dimensions (for $J_2>J_1$). Quantum fluctuations
induced by a transverse field are found to be unable to lift this degeneracy in favor of a classically
ordered state at the harmonic level. This remarkable fact suggests that a quantum-disordered ground state can be instead
promoted when non-linear fluctuations are accounted for, in agreement with existing results for
the isotropic case $J_1=J_2$. Moreover spin-wave theory shows
sizable regions of instability which are further candidates for quantum-disordered behavior.
\end{abstract}



\maketitle


\section{Introduction\label{introduction}}
Frustrated quantum magnets represent one of the richest playgrounds to
investigate quantum collective phenomena \cite{Milaetal2011}. Indeed
known models of frustrated quantum magnets might admit
quantum ground states without any classical counterpart, such as
valence bond crystals, or resonating-valence-bond spin liquids \cite{Balents10}.
Most of the investigations have focused on Heisenberg antiferromagnets,
due to their relevance to real compounds, although the lack of well
controlled analytical or numerical approaches for the bulk properties
of these systems still leaves the question open on the true nature of
their ground state \cite{Whiteetal11}.
On the other hand, quantum-disordered ground states have been shown to emerge
in models with strongly anisotropic interactions (in spin space and/or in real space)
which appear to provide the first controlled realizations of quantum spin liquids with
a topological nature \cite{Hermeleetal04, Isakovetal06, Banerjeeetal08, Isakovetal11}.
Some of these systems are related to quantum dimer models \cite{MoessnerQDM, Sikoraetal09},
which have provided the first known examples of quantum spin-liquid ground states
\cite{MoessnerS01-2}.

In this class of anisotropic systems a special role is played by frustrated quantum Ising models,
namely Ising systems enriched with quantum fluctuations, coming either from an exchange
coupling in the transverse spin components, or from a transverse magnetic field \cite{MoessnerS01, Chakrabartietal96}.
In the absence of quantum fluctuations, Ising models on frustrated lattices have generally
a \emph{classical} spin liquid nature, namely they exhibit an
exponential degeneracy of the ground state -- as it is the case for the Ising antiferromagnet on the
triangular lattice, kagom\'e lattice, checkerboard lattice, pyrochlore lattice, etc., with however
correlations that can be long-ranged, algebraic or short-ranged.
 The effect of quantum fluctuations is in general that of lifting the large degeneracy of
 ground states, leading either to the emergence of an ordered ground state
 (as \emph{e.g.} in the case of the triangular lattice \cite{MoessnerS01}) or to
 a ground state with novel spin-liquid properties (as it appears to be the case
 for the pyrochlore lattice \cite{Hermeleetal04, Banerjeeetal08}).
 In this study we focus on the Ising antiferromagnet on the checkerboard lattice, which represents
 a fundamental model of frustrated magnetism in two dimensions. Indeed, in the case
 of spatially isotropic interactions the ground state properties of the system are
 equivalent to those of the unbiased six-vertex model, also known as square ice
 \cite{Lieb67}. The effect of quantum fluctuations on such a system has been the
 subject of several recent investigations \cite{MoessnerS01, CastroNetoetal06,
 Moessneretal04, Shannonetal04, Zhou05, Ralkoetal10}, focusing particularly
 on the limit of weak quantum fluctuations, treated within degenerate perturbation
 theory.

 In the present study we adopt a different strategy, which allows us to treat arbitrarily
 strong quantum fluctuations in a generalized version of square ice. We investigate
 the $S=1/2$ Ising model on the checkerboard lattice with different couplings along the
 coordinate axes ($J_1$) and along the diagonals ($J_2$), as shown in Fig.~\ref{fig:lattice}.
 The anisotropy in the couplings allows to introduces a bias in the vertex weights of the corresponding
 vertex model, reducing the degeneracy to exponential in the linear dimensions of the
 system ($J_2 > J_1$), or even to a finite value (for $J_1 > J_2$). The application of a
 transverse field allows therefore to investigate the effect of quantum fluctuations on a
 classical manifold of states with variable degeneracy. Quantum fluctuations are
 investigated via linear spin-wave theory, accounting for the harmonic
 fluctuations around the classical ground state. Albeit limited to the harmonic approximation,
 such an approach allows for investigation of arbitrarily strong fields, and its breakdown signals
 the candidate regions in the phase diagram where novel quantum disordered behavior can be
 expected. On the side $J_2 > J_1$ the infinite
 degeneracy of the ground state would a priori make the spin-wave analysis impossible,
 given the exceedingly large number of possible classical reference states. In fact we
 demonstrate that the spin-wave spectrum does \emph{not} depend on the particular classical ground state
 chosen as a reference. This implies that the spectrum of linear excitations is defined unambiguously
 in the $J_2 > J_1$ region - this situation persists also in the isotropic case of square ice,
 $J_1 = J_2$, for which the lowest branch of the excitation spectrum is a flat band. But this
 implies as well that the classical degeneracy remains unaltered
 in presence of harmonic quantum fluctuations, which means that only non-linear quantum
 fluctuations can lift the degeneracy, a situation previously encountered in a number
 of frustrated magnets\cite{harris,chalker,ritchey,KorshunovDoucotPhysRevLett.93.097003,Korshunov05,henley,coletta}. Moreover, harmonic quantum fluctuations triggered by a sufficiently
 strong field are found to reverse the classical
 hierarchy of ordered states close to the isotropic ($J_1=J_2$) limit, suggesting
 that anharmonic fluctuations might destabilize the classical order completely.
 Finally the classical order parameter is found to be completely washed out by
 quantum fluctuations for strong fields close to the classical polarization transition,
 and for couplings close to the isotropic case.  We can therefore
 conclude that the transverse field Ising model (TFIM) on the checkerboard lattice can
 harbor quantum disordered phases for strong frustration and fields, and that non-linear
 quantum fluctuations are expected to play a major role in the case of extensive
 degeneracy of the ground state.

 The paper is structured as follows: Sec.~\ref{latticeham} describes the model, its
 behavior in zero applied field, and the classical behavior in a transverse field;
 Sec.~\ref{perturbation} reviews known results on quantum square ice from degenerate
 perturbation theory and numerics; Sec.~\ref{lsw} describes linear spin-wave theory as
 applied to the various regimes of the checkerboard lattice TFIM; Sec.~\ref{results}
 discusses the phase diagram emerging from spin-wave theory; and finally conclusions
 are drawn in Sec.~\ref{conclusions}, along with a discussion about physical realizations.
 The general framework of spin-wave theory for
 the TFIM on arbitrary lattices is presented in Appendix~\ref{diag}, while the spin-wave
 observables for the checkerboard lattice are discussed in Appendix~\ref{diagNc}.

\section{Classical behavior of the trasverse-field Ising model on the checkerboard lattice
\label{latticeham}}
\subsection{Model Hamiltonian and ground-state properties in  zero field}
\label{notations}
\begin{figure}
\epsfig{file=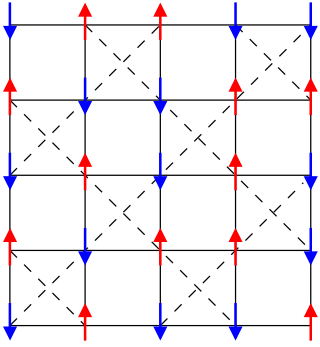,scale=1.}
\caption{Checkerboard lattice. Full lines represent couplings with strength
 $J_{1}$, dashed lines represent couplings with strength $J_{2}$. }
\label{fig:lattice}
\end{figure}

 The Hamiltonian of the TFIM on the checkerboard lattice reads
 \begin{equation}
\label{eq:ham}
	\mathcal{H}=		J_1\Sum_{\langle i,j \rangle}{S_i^zS_j^z}
							+J_2\Sum_{\langle\langle i,j \rangle\rangle}{S_i^zS_j^z}
							-\Gamma\Sum_{i}{S_i^x}~.
\end{equation}
where ${\bm S}_i$ are quantum spins of length $S$, satisfying $|{\bm S}_i|^2 = S(S+1)$
and $[S_i^{\alpha}, S_i^{\beta}] = i\varepsilon_{\alpha\beta\gamma} S_i^{\gamma}$.
The first sum in Eq.~\eqref{eq:ham} runs over the nearest-neighbor bonds of a square lattice,
while the second sum runs over the next-to-nearest-neighbor (diagonal) bonds
on a staggered array of plaquettes (see Fig.~\ref{fig:lattice}). We consider
here frustrated antiferromagnetic couplings $J_1$, $J_2 > 0$.
$\Gamma$
is a transverse magnetic field, introducing quantum fluctuations in the system.
As we will see in the following section, in zero field the ground-state
properties of the above model are equivalent to those of an $m$-vertex model with $m=2$, 4 or
6 depending on the Hamiltonian parameters. Motivated by this equivalence, in the following
we will indicate as \emph{vertices} (denoted by $\boxtimes$) the squares with additional
diagonal $J_2$ couplings, and as \emph{plaquettes} (denoted by $\square$) the
squares without diagonal couplings.

 When $\Gamma = 0$ we can easily rewrite the Hamiltonian in the following form
 \begin{equation}
 {\cal H} = J_2~ {h}_{\rm ice}
 + (J_1-J_2) \sum_{\langle i,j \rangle} S_i^z S_j^z
 \label{eq:J1_J2}
 \end{equation}
 if $J_1 >  J_2 $, and
 \begin{equation}
 {\cal H} = J_1 ~{h}_{\rm ice}
 + (J_2-J_1) \sum_{\langle\langle i,j \rangle\rangle} S_i^z S_j^z
 \label{eq:J2_J1}
 \end{equation}
  if $J_2 > J_1$, where we have introduced the square-ice Hamiltonian
  \begin{equation}
  {h}_{\rm ice} = \sum_{\boxtimes} \left[  \left(\sum_{i\in \boxtimes} S_i^z\right)^2 - 4\sum_{i\in\boxtimes}(S_i^z)^2 \right]~.
  \end{equation}
  With this choice the ground state is identified in two steps:
 1) firstly, one needs to impose on each spin the constraint that $S_i^z = \pm S$,
and on each vertex the zero-magnetization constraint,
 $M_{\boxtimes} = \sum_{i\in \boxtimes} S_i^z = 0$: these two constraints minimize the first term on
 the right hand side of both Eq.~\eqref{eq:J1_J2} and \eqref{eq:J2_J1}.
 The $M_{\boxtimes}=0$ constraint corresponds to the so-called \emph{ice rule} for square ice, and therefore
 we will hereafter denote the states which satisfy it (and which satisfy $S_i^z = \pm S$) as ice-rule states;
 2) secondly, one needs to
 choose, among the ice-rule states, those corresponding to a minimum of the second term
 in the r.h.s. of Eqs.~\eqref{eq:J1_J2} and \eqref{eq:J2_J1}; this term is always
 antiferromagnetic by construction, so that it can be minimized by zero-magnetization vertices.

\begin{figure}
\epsfig{file=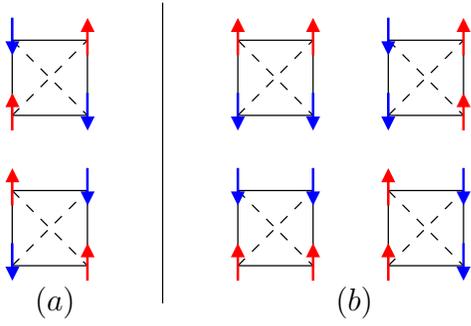,scale=1.}
\caption{Crossed plaquettes that obey the ``ice rule''. (a) On the N\'eel plaquettes, the nearest
neighbor links are satisfied. (b) On the ``collinear'' plaquettes, the next nearest neighbor links are
satisfied.}
\label{fig:vtype}
\end{figure}

In the special case $J_1=J_2$ we recover ${\cal H} = J_{1(2)} h_{\rm ice}$,
namely the square ice model, whose ground states are only constrained by the
ice rule (along with the condition $S_i^z = \pm S$). The ground state properties of this
system are equivalent to the 6-vertex model, displaying an exponential degeneracy of the
ground state \cite{Lieb67}. This degeneracy is (partially) lifted when $J_1 \neq J_2$.

If $J_1 > J_2$ we need to select among all zero-magnetization vertices
with satisfied antiferromagnetic $J_1$ bonds. This further constraint imposes that each vertex
must take the antiferromagnetic (N\'eel) configuration in Fig.~\ref{fig:vtype}(a) or its
spin-flipped version (2-vertex model), and it reduces the possible ground states to the
two N\'eel-ordered states. The ground-state energy per spin takes the value
$E_{\text{N\'eel}} = - (2J_1-J_2)S^2$.

If on the contrary $J_2 > J_1$, the zero-magnetization vertices
must satisfy the $J_2$ bonds, and therefore they take one of the
4 collinear configurations in Fig.~\ref{fig:vtype}(a), reducing the ground state
of the system to that of a 4-vertex model. The collinear ground states
(namely containing only collinear
vertices) are massively degenerate, because flipping a linear chain of spins
along the $J_2$ diagonals does not alter any of the constraints which the
ground state must satisfy. Therefore the ground-state degeneracy is equal
to $2^{N_d}$, where $N_d$ is the number of $J_2$ diagonals.
The ground-state energy per spin takes the value
$E_{\rm coll} = - J_2 S^2$.

 In summary the degeneracy of the ice-rule states is partially lifted when  $J_1 \neq J_2$,
 and ice-rule states are organized into a band with energy width (per spin) given by
\begin{equation}
\omega  = |E_{\text{N\'eel}} - E_{\rm coll} | = 2 |J_1 - J_2| S^2~.
\end{equation}

The above discussion is valid for an arbitrary value of the spin length $S$,
and in particular also in the classical limit $S\to\infty$, in which one can
introduce continuous spins $\tilde{\bm S}_i = {\bm S_i}/S$ with unit
length $|\tilde{\bm S}_i|^2 =1$. On the other hand, we will see that
the application of a transverse field introduces significant differences
for the ground-state properties depending on the spin length $S$.

\subsection{Excited states  in zero field}

 The nature of the excited states is dependent upon the spin length.
 In the following we will concentrate on the $S=1/2$ case, which will be the main
 focus of this paper. In this limit, when $J_1 > J_2$ the lowest energy
 excitations correspond either to:

 1) a single spin flip, costing an energy
 \begin{equation}
 \Delta =  (2-\nu_1) J_1 + (1-\nu_2)J_2
 \end{equation}
 where $\nu_{p}$ is the number of frustrated $J_{p}$ links ($p=1,2$) connected to each site.
For N\'eel states, $\nu_1 = 0$ and $\nu_2 = 2$, while for collinear states
$\nu_1 = 2$ and $\nu_2 = 0$.
Notice that for generic ice-rule states one has $\nu_{1}+\nu_{2}=2$.

\begin{figure}[h!]
\epsfig{file=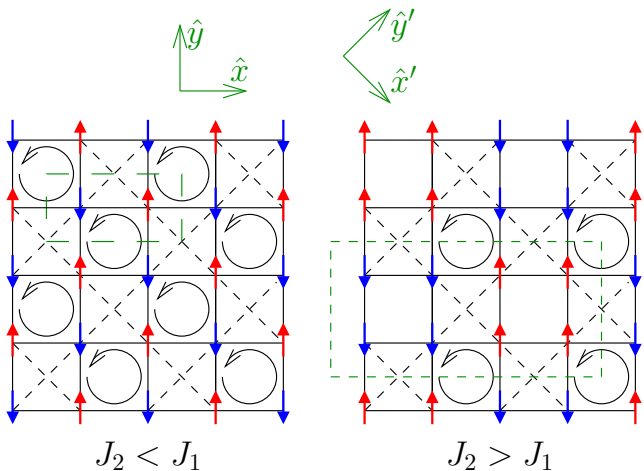,scale=.9}
\caption{Reference classical ground states. The circles indicate flippable plaquettes.}
\label{fig:classicalgs}
\end{figure}

2) a joint flip of all the spins on a plaquette (plaquette flip). This operation
has the lowest energy when applied to a \emph{flippable} plaquette,
with the property that its flip connects the initial ice-rule state to another
ice-rule state. All the neighboring vertices of a flippable plaquette share
with the plaquette a bond with zero magnetization, so that the plaquette flip will
 not alter the vertex magnetization. This imposes that the flippable plaquette has
a local N\'eel configuration. In N\'eel states all plaquettes are flippable,
with an energy cost of $\Delta_{\rm plaq} = 4(J_1 - J_2)$.
In the collinear states, only a portion of the plaquettes are flippable
(at most one half as in the state depicted in Fig.~\ref{fig:classicalgs}), with an energy
cost of $\Delta_{\rm plaq} = 4(J_2 - J_1)$.

Comparing the energy cost of a plaquette flip with that of a spin flip, we find
that plaquette flips are the lowest-energy excitations in the
parameter range $2/3 < J_2/J_1 < 4/3$, and outside of this range
spin flips are instead the excitations with the lowest energy.

\subsection{$S\to\infty$ limit in a transverse magnetic field
\label{clrot}}

 In this section we discuss the effect of a transverse field on classical continuous
 spins ($S\to\infty$). The corresponding ground state configurations will serve
 as a template for the spin-wave analysis in the quantum case.
 The transverse field introduces a canting of the spins along the $x$ axis
 by an angle $\vartheta$. This angle increases with the intensity of
the field up to a critical value $\Gamma_{c}$ at which $\vartheta = \pi/2$,
corresponding to the polarization of the spins along $x$.
For both the collinear and the N\'eel ground states, each spin sees the
same local field (in modulus) created by the neighboring spins
(${\bm h}_{\rm loc} = \pm (4J_1 - 2J_2)S {\hat z}$ for N\'eel states,
and  ${\bm h}_{\rm loc} = \pm 2J_2 S {\hat z}$ for collinear states).
Therefore the application of the external field will create the same canting
angle $\vartheta$ on each spin.  Upon canting, the spin configuration
becomes $S^{x}_{i}=S\sin\vartheta$ and $S^{z}_{i}
=\pm S\cos\vartheta$.

\begin{figure}[h]
\epsfig{file=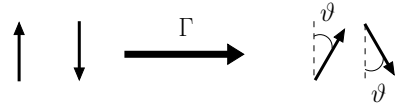,scale=.4}
\caption{Classical rotation angle induced by the transverse field.}
\label{fig:rotation}
\end{figure}

The classical energy per spin admits the compact expression, valid for both
the N\'eel and collinear states:
\begin{equation}
\label{clen}
\begin{array}{lcl}
\varepsilon_{cl}&=& \left[(\nu_{1}-2)J_{1} + (\nu_{2}-1)J_{2}\right]S^{2}\cos^{2}\vartheta - \Gamma S \sin\vartheta\\
					&=& \left[\nu_{2}\left(J_{2}-J_{1}\right)-J_{2}\right]S^{2}\cos^{2}\vartheta - \Gamma S\sin\vartheta
\end{array}
\end{equation}

\begin{figure}[h!]
\begin{center}
\rotatebox{270}{
\epsfig{file=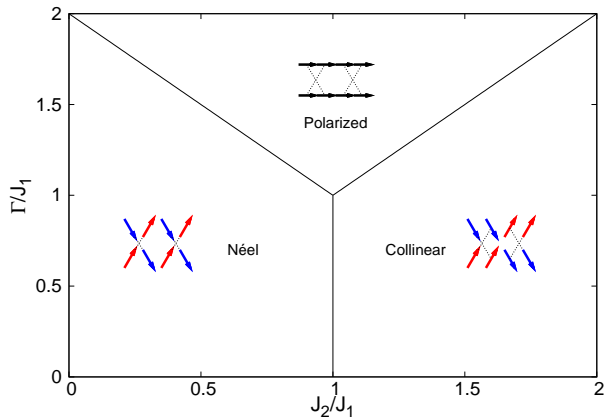,scale=0.32}
}
\caption{Classical phase diagram of the transverse-field Ising model on the checkerboard lattice.}
\label{fig:clpd}
\end{center}
\end{figure}

Minimizing the energy per spin with respect to $\vartheta$, we find
\begin{equation}
	\sin\vartheta = \text{min}\left(\dfrac{\Gamma}{2S\left[J_{2}-\nu_{2}\left(J_{2}-J_{1}\right)\right]},1\right)
\end{equation}
If $\Gamma> \Gamma_c = 2S\left[J_{2}-\nu_{2}\left(J_{2}-J_{1}\right)\right]$, the system
becomes completely polarized in the transverse direction. 
For $\Gamma> \Gamma_c$ the classical ground-state
energy per spin takes the value
\begin{equation}
\label{clenergy}
	\varepsilon_{cl}=-S^2\left[J_{2}-\nu_{2}\left(J_{2}-J_{1}\right)\right] \left(1+\sin^{2}\vartheta\right)~.
\end{equation}
The resulting classical phase diagram for $S\to\infty$ is shown in Fig.~\ref{fig:clpd}.



\section{Square ice in a transverse field : results from perturbation theory}
\label{perturbation}

For square ice ($J_1 = J_2$),
in the case $\Gamma \ll J_1, J_2$ the transverse field can be treated via
degenerate perturbation theory on the manifold of ice-rule states.
Considering all terms up to fourth order \cite{MuellerHartmannR02}, one can easily find the
following effective Hamiltonian for the subspace of ice-rule states (up to
an additive constant):
\begin{equation}
	\mathcal{H}_{\rm eff}=  -\dfrac{\Gamma^{4}}{\Delta^{3}}\Sum_{{\rm flippable}~ \square}
	\left({S^{+}_{i}S^{-}_{j}S^{+}_{k}S^{-}_{l}}  + {\rm h.c.}\right) + {\cal O}\left(\frac{\Gamma^6}{\Delta^5}\right)
	\label{eq:Heff}
\end{equation}
where $(ijkl)$ are the four sites on a plaquette (in clockwise order), and the sum runs on flippable
plaquettes. Various claims exist in the
literature \cite{MoessnerS01, CastroNetoetal06} that this Hamiltonian will lift
the degeneracy of ice-rule states in favor of the N\'eel state, based on the
fact that the N\'eel state has the largest number of flippable plaquettes.
In fact, this Hamiltonian is completely off-diagonal for the ice-rule states, so
that it cannot favor a specific ice-rule state, but only a resonant superposition
thereof. In particular a flippable plaquette, once flipped, turns its four neighboring
vertices from N\'eel configurations to collinear configurations or viceversa, namely
the effective Hamiltonian will resonantly connect local N\'eel configurations
with local collinear ones. At the same time, two corner-sharing flippable plaquettes
cannot be both flipped without leaving the ice-rule manifold, which suggests
that N\'eel-collinear resonances will be localized to single plaquettes.
A rough approximation to the ground state of Eq.~\eqref{eq:Heff} is
therefore a state in which a checkerboard subset of non-corner sharing plaquettes
resonate between two flippable states, giving rise to a resonating-plaquette solid (RPS)
\begin{equation}
|\Psi_0\rangle \approx |\Psi_{\rm RPS}\rangle =
{\prod_{\square}}' \left(|\uparrow_i \downarrow_j \uparrow_k \downarrow_l \rangle +
|\downarrow_i \uparrow_j \downarrow_k \uparrow_l \rangle\right)/\sqrt{2}
\end{equation}
where the primed product runs on a sublattice of plaquettes. This state breaks
the two-fold symmetry between the two plaquette sublattices.
While direct numerical investigations of the square ice model in a transverse
field are not known to us, there exists in the literature a series of
numerical studies of square ice with different quantum perturbations,
which all map perturbatively onto the effective Hamiltonian of Eq.~\eqref{eq:Heff}.
A direct simulation of Eq.~\eqref{eq:Heff} is reported in Ref.~\onlinecite{Moessneretal04},
while Refs.~\onlinecite{Shannonetal04}, \onlinecite{Zhou05} and  \onlinecite{Ralkoetal10}
focus on hardcore bosons on the checkerboard lattice
at half filling, with  strong nearest-neighbor repulsion, and
weak hopping (with either positive or negative sign) between nearest
neighbors and next-to-nearest neighbors. The latter model is equivalent to square ice
with weak ferromagnetic/antiferromagnetic exchange terms for the $x$ and $y$
spin components. The common result of all these studies is that
the ground state for weak quantum perturbations has indeed long-range
RPS order, and no magnetic order. Therefore we expect spin-wave theory
to break down or become inconclusive in this limit -- which is indeed one of the main
results of the following analysis.

\section{Linear Spin-Wave theory
\label{lsw}}
In the following we describe a treatment of quantum fluctuations introduced by the transverse
field in the $S=1/2$ case based on a linear spin wave expansion\cite{misguich,coletta}. We will then treat
separately the spectrum of excitations above the various classical reference states
of the system: N\'eel, collinear, and fully polarized.

\subsection{Spin-boson transformation
\label{lswham}}
We begin by considering a generic classical ground state with long-range magnetic
order, and with a magnetic unit cell containing $n$ spins. We denote $S_{l,p}$ the {\it p}-th spin ($p=1\ldots n$) of the {\it l}-th cell.
As seen in Section \ref{clrot}, in the classical limit an applied transverse field rotates
the spins around the $y$-axis by an angle $\vartheta$. We introduce a local rotation of the
spin configuration, $\tilde{{\bm S}}_{l,p} = \sigma_p \mathcal{R}_y (\sigma_{p}\vartheta) {\bm S}_{l,p}$,
where $\sigma_p = 1(-1)$ if the spin in zero field has positive (negative) projection along  the
$z$ axis, and $\mathcal{R}_y(\pm \vartheta)$ is the rotation matrix of an angle $\pm \vartheta$
around the $y$ axis. In the classical limit $S\to\infty$ the ground state is a simple ferromagnetic
state for the $\tilde{{\bm S}}_{l,p}$ spins, namely $\tilde{S}^{z}_{l,p} = S$ everywhere.

We then consider small quantum fluctuations around this classical reference state, by
transforming the quantum spins to bosons via a linearised Holstein-Primakoff transformation
\cite{HP} valid in the limit of a small number of bosons $n_{l,p} \ll 2S$:
\begin{equation}
\tilde{S}^{z}_{l,p} = S-a_{l,p}^{\dagger}a_{l,p} ~~~~~ \tilde{S}^{x}_{l,p} \approx \sqrt{\dfrac{S}{2}}\left(a_{l,p}^{\dagger}+a_{l,p}\right)	~.	
\end{equation}
Here $a_{l,p}$ and $a_{l,p}^{\dagger}$ are bosonic operators, satisfying $[a_{l,p},a_{l,p}^{\dagger}]=1$ and
$[a_{l,p}^{(\dagger)},a_{l,p}^{(\dagger)}]=0$.

\subsection{Harmonic Hamiltonian for ordered ice-rule states
\label{lswham2}}

The Hamiltonian is then expanded up to quadratic order in the bosonic operators (the linear terms vanish
by construction). In the following we will specialize the discussion to reference classical states
which in zero field are ice-rule states with long-range magnetic order, namely the states
which minimize the energy in the classical limit $S\to\infty$, and whose ordered structure
allows to build a spin-wave theory. These states have the property that the number of
frustrated bonds of type 1 and 2, $\nu_1$ and $\nu_2$, is the same for every site. Under these
generic assumptions the quadratic bosonic Hamiltonian reads
\begin{equation}
	\label{eq:lswham}
	\mathcal{H}_{\rm LSW}= N\varepsilon_{\rm cl} + J_{1}\tilde{\mathcal{H}}_{nn} + J_{2}\tilde{\mathcal{H}}_{nnn} + \Gamma\tilde{\mathcal{H}}_{\Gamma}
\end{equation}
with
\begin{widetext}
	\begin{equation}
	\label{reallsw}
		\begin{array}{lcl}
			\tilde{\mathcal{H}}_{nn}&=&2S~\nu_{2}\cos^{2}\vartheta\Sum_{l,p}{a_{l,p}^{\dagger}a_{l,p}}
			+\dfrac{S}{2}\sin^{2}\vartheta\Sum_{\langle lp,l'p'\rangle}{\left(a_{l,p}^{\dagger}a_{l',p'}^{\dagger}
			+a_{l,p}^{\dagger}a_{l',p'}+h.c.\right)}\\
			\tilde{\mathcal{H}}_{nnn}&=&2S(1-\nu_{2})\cos^{2}\vartheta\Sum_{l,p}{a_{l,p}^{\dagger}a_{l,p}}
			+\dfrac{S}{2}\sin^{2}\vartheta\Sum_{\langle\langle lp,l'p'\rangle\rangle}{\left(a_{l,p}^{\dagger}a_{l',p'}^{\dagger}
			+a_{l,p}^{\dagger}a_{l',p'}+h.c.\right)}\\
			\tilde{\mathcal{H}}_{\Gamma}&=& -\Gamma\sin\vartheta\Sum_{l,p}{a_{l,p}^{\dagger}a_{l,p}}
		\end{array}
\end{equation}
\end{widetext}
This is a remarkable result, in that the spin-wave Hamiltonian depends only on the
frustration parameters $\nu_1$ and $\nu_2$, while it is completely independent of the geometry of the unit cell. The frustration
parameters  $\nu_1, \nu_2$ distinguish among N\'eel states and collinear states, but they are \emph{not}
able to distinguish among different collinear states. At the square-ice point $J_1=J_2$ the dependence
on the frustration parameter drops. Therefore, as we will discuss further,
quantum corrections at the harmonic level are not able to lift the degeneracy between ordered
ice-rule states, regardless of the size of their magnetic unit cell. This result can be extended
even to disordered ice-rule states, which can be regarded as ordered ones with an infinite unit cell.

To diagonalize the spin-wave Hamiltonian, Eq.~\eqref{reallsw},
we first introduce the Fourier transform of the bosonic operators
and then perform a $n$-modes Bogoliubov transformation as described in
Appendix \ref{diag}. The Hamiltonian then becomes
\begin{equation}
\label{rotham}
\mathcal{H} = N\left(\varepsilon_{cl}-\dfrac{\varepsilon_{0}}{2}\right)
					 +\dfrac{1}{2}\Sum_{\textbf{k},p}{\omega_{\textbf{k},p}\left(b_{\textbf{k},p}^{\dagger}b_{\textbf{k},p}+\dfrac{1}{2}\right)}
\end{equation}
 where $\varepsilon_{0}=2S\left[J_{2}-\nu_{2}\left(J_{2}-J_{1}\right)\right]$~.

\subsection{N\'eel state
\label{neel}}
\begin{figure}
\begin{center}
	\includegraphics[width=9cm]{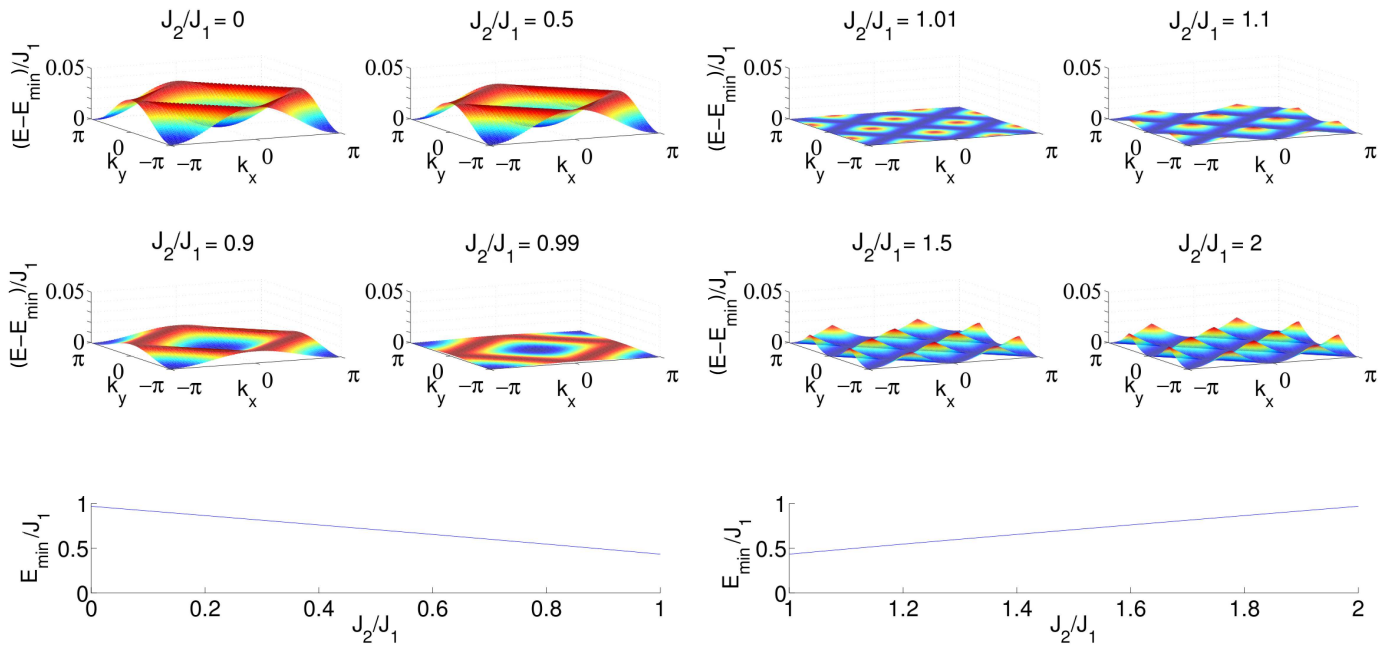}
	\caption{Lowest band of the magnon spectrum for $\Gamma=J_{1}/2$ and various
	values of $J_{2}/J_{1}$, around the N\'eel state. For the purpose of readability, the bands
	have been offset by the energy of their lower edge. The lower band edge (corresponding
	to the minimum excitation gap) is plotted in the lower panel.}
	\label{fig:spN}
\end{center}
\end{figure}

Let us first consider the N\'eel state, defined for $\Gamma<2S(2J_{1}-J_{2})$.
Its unit cell contains $n=2$ spins (see figure \ref{fig:classicalgs}). The unit cells
form a rotated square lattice with vectors $\hat{x}'=\left(\hat{x}-\hat{y}\right)$
and $\hat{y}'=\left(\hat{x}+\hat{y}\right)$.
The diagonalization of $M_{\kb}$ shows that the spectrum of the magnon excitations
is gapped whenever the classical N\'eel state is defined - {\sl i.e.} if $\Gamma<2S(2J_{1}-J_{2})$.
Its lower band is plotted in figure \ref{fig:spN}. It has minima at $(0,0)$ and at the four corners
of the Brillouin zone. This corresponds to the structure of the classical N\'eel state.

\subsection{Collinear states
\label{collinear}}

 As already mentioned in Sec.~\ref{lswham}, all collinear states admit the same
 frustration parameter $\nu_2$, and hence the same spin-wave Hamiltonian.
 This means that they possess the same spectrum of harmonic spin-wave excitations
 (but folded into a smaller Brillouin zone, the larger the unit cell),
 and that zero-point quantum fluctuations cannot lift the degeneracy among them.

\begin{figure}
\begin{center}
	\includegraphics[width=9cm]{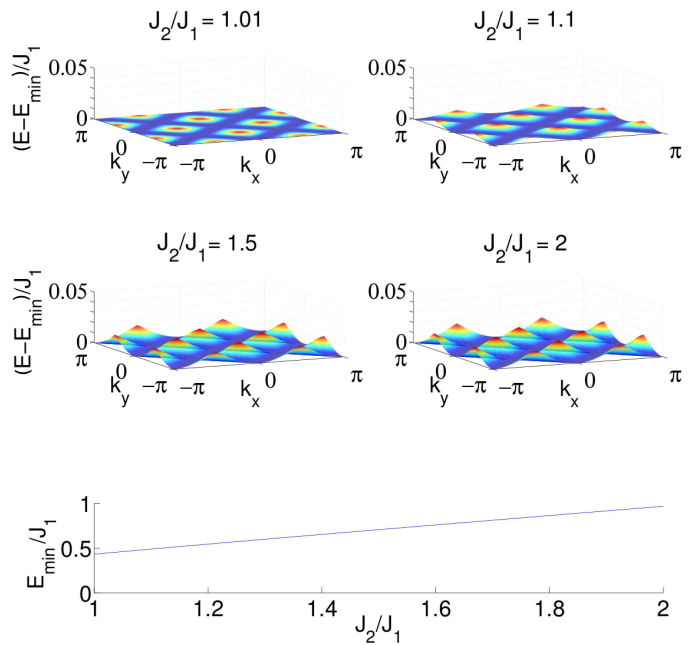}
	\caption{Lowest band of the magnon spectrum for $\Gamma=J_{1}/2$ and various
	values of $J_{2}/J_{1}$, around any collinear state. For the purpose of readability, the bands
	have been offset by the energy of their lower edge. The lower band edge (corresponding
	to the minimum excitation gap) is plotted in the lower panel.}
	\label{fig:spC}
\end{center}
\end{figure}

We will then specify the discussion to the particular collinear state represented in
figure \ref{fig:classicalgs}. Its unit cell contains $n=8$ spins; the unit cells form a square lattice
with vectors $\hat{x}'=2\left(\hat{x} -\hat{y}\right)$ and
$\hat{y}'=2\left(\hat{x}+\hat{y}\right)$.
While not being the simplest of all collinear states,
this state is relevant because it can be energetically stabilized against other
collinear states by \emph{e.g.} dipolar interactions, which are relevant for realistic
ice models \cite{Henryprep}.
The magnon dispersion relation, obtained by diagonalizing $M_{\kb}$,
is shown in Fig.~\ref{fig:spC}. It shows a finite gap, and two \emph{lines}
of minimum-energy degenerate modes along the axes of the first
Brillouin zone of the magnetic lattice (1/8 of the Brillouin zone of
the geometric lattice, shown in Fig.~\ref{fig:spC}). These degenerate modes
traveling with momentum $(k_x, \pm k_x)$ for all $k_x$ values can be
associated with deconfined monopole-like\cite{CMS} pairs, obtained by flipping a finite string
of spins along a $J_2$-diagonal of the checkerboard lattice. Given the degeneracy
of all collinear states, not perturbed by quantum fluctuations, these pairs are
deconfined along the $J_2$-diagonals, and their energy is independent
of momentum as long as it satisfies the constraint of diagonal motion.

\subsection{Quantum square ice
\label{dege}}

At the square-ice point $J_1=J_2=J$, spin-wave theory built around any ice-rule state
with canting spins produces the same excitation spectrum and zero-point quantum fluctuations.
Therefore harmonic quantum fluctuations are not able to lift the degeneracy of the classical
ice-rule manifold, and the elementary excitations are identical to the classical case,
namely deconfined monopole pairs moving with arbitrary momentum. This is reflected
in the spin-wave dispersion, showing a \emph{perfectly flat} band for the lowest energy
excitations, and with a gap equal to the classical value, namely $\Delta_{cl}=2\varepsilon_{0}=2JS^{2}$.

\subsection{Polarized states
\label{polarized}}
For large $\Gamma$, the classical reference state is totally polarized along the field.
Nonetheless quantum fluctuations introduce deviations from the polarized state,
given that the field term in the Hamiltonian does not commute with the Ising
couplings. We build a spin-wave Hamiltonian around the polarized state analogously
to what has been done for the ice-rule states. Even though in the classical spin configuration
all the spins have the same orientation, the magnetic unit cell contains two
sites (exactly as in the case of the N\'eel state), due to the fact that the checkerboard
lattice is not a Bravais lattice. The bosonic excitations correspond to deviations of
the spin from full polarization along the $x$ axis.

\begin{figure}
\begin{center}
	\includegraphics[width=9cm]{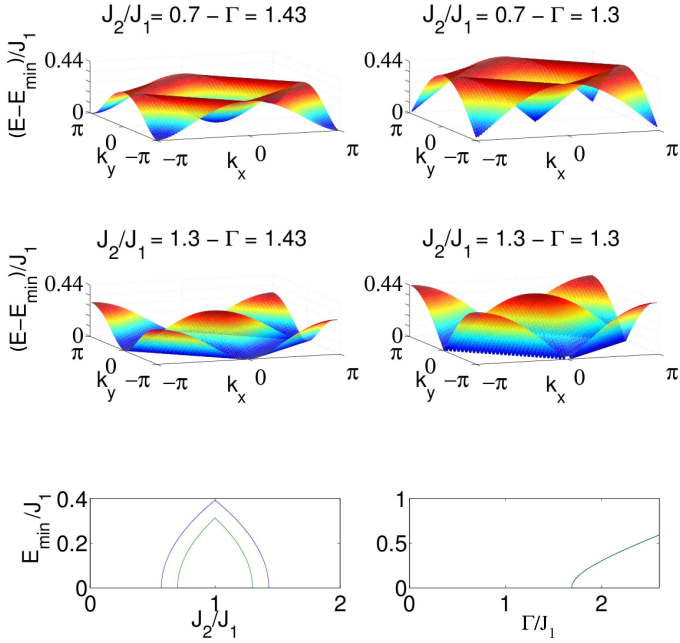}
	\caption{Upper panels: lowest band of the magnon spectrum around
	the polarized state for $\Gamma > \Gamma_{c}$ (left column) and
	for $\Gamma = \Gamma_{c}$ (right column). Lower panels:
	Minimum excitation gap: (left) as a function of $J_{2}/J_{1}$ for
	$\Gamma/J_{1}=1.3$ (green line) and for $\Gamma/J_{1}=1.43$
	(blue line); (right) as a function of $\Gamma/J_{1}$ for $J_{2}/J_{1}=0.7$
	and $1.3$ (the curves for both cases coincide).}
	\label{fig:spP}
\end{center}
\end{figure}

The magnon spectrum is shown in Fig.~\ref{fig:spP}. It displays softer modes at
the four corners of the square-lattice Brillouin zone (for $J_1 > J_2$), and along the edges
of the checkerboard-lattice Brillouin zone (for $J_1 < J_2$). These modes become gapless
when approaching the critical field $\Gamma_c$, signaling the instability of the
fully polarized state to a N\'eel state (for $J_1 > J_2$) and to degenerate collinear states (for $J_1 < J_2$).


\section{Results of the spin wave analysis
\label{results}}
The main observables from linear spin-wave theory are represented by
the internal energy $E = \langle {\cal H}_{\rm LSW}\rangle$ and the order parameter
$m = \langle {\tilde S}^z \rangle/S$, whose expressions are given in Appendix
\ref{diagNc}. These two quantities allow us to extract the quantum phase diagram of our
system in the harmonic approximation.

\begin{figure}[htpb!]
\begin{center}
\rotatebox{270}{
\epsfig{file=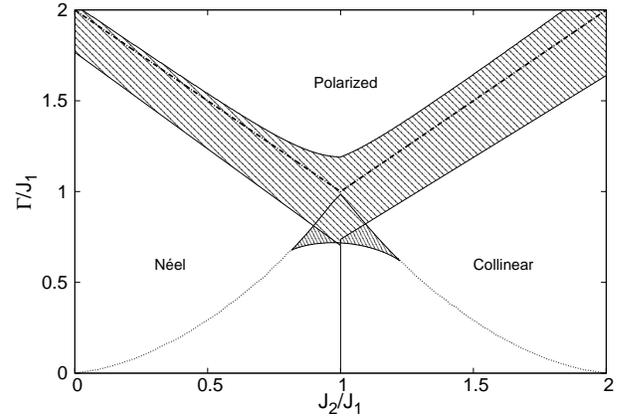,scale=0.32}
}
\caption{Phase diagram of the $S=1/2$ transverse-field Ising model on a checkerboard
lattice from spin-wave theory. The hatched, and dense-hatched regions correspond
to the next-to-nearest neighbor correlation inversion region and to the energy-hierarchy inversion region,
respectively (see text). Below the dotted lines one can find - in the classical
limit - two states which are (local) energy minima and which have order in the $z$ spin components:
one state with N\'eel order, and the other with collinear order. Above this line one of the two states
becomes the polarized state. Therefore collinear and N\'eel ordered states can only be compared
energetically below the dotted line.}
\label{fig:pd}
\end{center}
\end{figure}

\begin{figure}[htpb!]
\begin{center}
\rotatebox{270}{
\epsfig{file=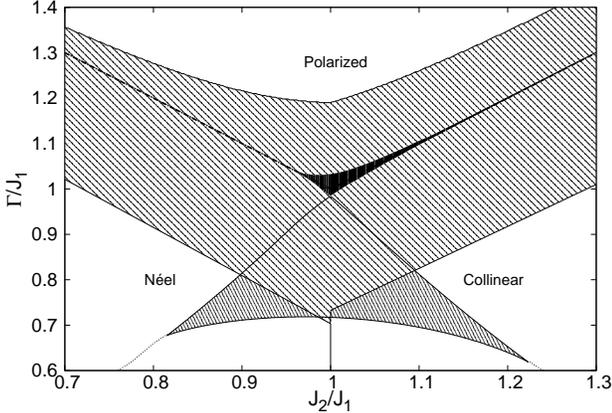,scale=0.32}
}
\caption{Zoom of Fig.~\ref{fig:pd} around the classical tri-critical point.
In the black domain, the order parameters of all considered phases vanish.
Other marked zones are as
in Fig.~\ref{fig:pd}.}
\label{fig:pdz}
\end{center}
\end{figure}

For each value of $J_{2}$ and $\Gamma$, we can a priori choose between three families
of reference states as candidate ground state: the N\'eel states, the fully polarized state and
the degenerate collinear states. Notice that classically, the N\'eel and the collinear states differ
in the value of the correlations between next-to-nearest neighbors $C^{(2)}=
\langle S_{i}^{z}S_{j}^{z}\rangle_{\langle\langle i,j\rangle\rangle}=(\nu_{2}-1)$.
For the N\'eel state, $C^{(2)}_{\text{N\'eel}}=S^{2}$, whereas for all
collinear states  $C^{(2)}_{\text{collinear}}=-S^{2}$.
A natural definition of the ground state (or ground-state
manifold) identified by spin wave theory is the state (or the family of states) which
has the lowest energy, whose order parameter is finite and whose next-to-nearest-neighbor
correlations have the correct sign; moreover the stability of the state requires also that the
spin-wave frequencies be real numbers.
The satisfaction of these four conditions allows us to identify the phases indicated
in Figs.~\ref{fig:pd} and ~\ref{fig:pdz}, which correspond to the classical phases for the same parameter
ranges.

\begin{figure}
\begin{center}
\rotatebox{270}{
\epsfig{file=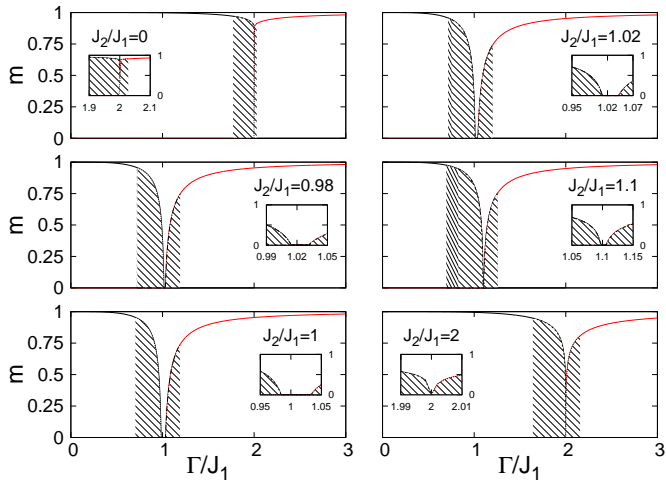,scale=0.32}
}
\null\vspace*{.5cm}
\caption{Order parameter vs field for different values of $J_{2}$. It vanishes at the critical
value of the field and a small gap opens for $J_{2}$ close to $J_{1}$ where none of the
three states has a finite order parameter. The insets are closer views around the critical value
of the field. The hatched and dense-hatched regions correspond those identified in the
phase diagram (Fig.~\ref{fig:pd}).}
\label{fig:op}
\end{center}
\end{figure}

A clear region of instability of spin-wave theory is found close to the classical
transition line between the fully polarized phase and the N\'eel and collinear phases, as
indicated by the black region in Fig.~\ref{fig:pdz}. In this parameter range the
order parameters for \emph{all} reference states are found to vanish, as shown in
Fig.~\ref{fig:op}, clearly signaling the onset of a quantum-disordered phase.

However, a number of other anomalies revealed by the linear-spin wave theory close
to the transition to the paramagnetic phase point to a significantly larger region
where the classical behaviour is probably destroyed by quantum fluctuations.

\begin{figure}
\rotatebox{270}{
\epsfig{file=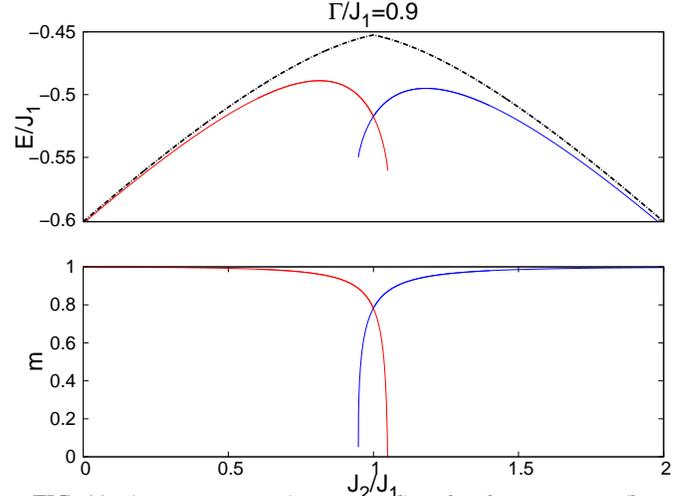,scale=0.32}
}
\caption{Average energy (upper panel) and order parameter (lower panel) associated
with the N\'eel (red solid line) and collinear (blue solid line) reference states. The
dashed lines correspond to the classical energies. Both panels are for $\Gamma/J_{1}=0.9	$.}
\label{fig:observables}
\end{figure}

First of all, a qualitative deviation from the classical behavior is observed when
approaching the square ice limit $J_1 = J_2$ in a strong field $0.7 \lesssim \Gamma \lesssim J_1$.
In this range (indicated by the dense-hatched region in Fig.~\ref{fig:pd}) we observe an inversion of the
energy hierarchy between N\'eel and collinear states with respect to the classical case. This occurs despite the fact that quantum fluctuations
are stronger for the energetically favored phase, as shown by the order parameters of the two
phases obeying an opposite hierarchy (namely $m_{\text{N\'eel}} > m_{\rm collinear}$ when
$E_{\text{N\'eel}} > E_{\rm collinear}$, and viceversa, see Fig.~\ref{fig:observables}).
The inversion in the energy hierarchy is due to strong quantum corrections to the classical
energy, which change qualitatively the dependence of both N\'eel and collinear energies on $J_2/J_1$.

This strong quantum effect of energy hierarchy inversion suggests that
classical order might be unstable around the hierarchy inversion region in Fig.~\ref{fig:pd}
when considering quantum fluctuations beyond linear spin-wave theory. The real ground state
of the system may then be an intermediate phase which cannot be described within the linear
spin wave approximation.

Another strong quantum effect is revealed close to the classical phase boundaries.
 While classically $E_{\text{N\'eel}}$  and $E_{\rm collinear}$ are monotonic functions of $J_2/J_1$,
they become non-monotonic around the above mentioned field range. In particular $E_{\text{N\'eel}}$ grows with
increasing $J_2/J_1$ until it reaches a maximum, beyond which it starts to decrease; from a classical
point of view this is quite surprising. According to the Hellmann-Feynman theorem, the next nearest neighbor
correlations $C^{(2)}$ are given by the derivative of the energy with respect to $J_{2}$, namely
$C^{(2)}=\partial{\langle \mathcal{H}\rangle}/\partial{J_{2}}$. Consequently, a change of sign in the derivative of $E$
corresponds to a change of sign in $C^{(2)}$, which means that the harmonic ground state is dramatically different from
the reference state. The locus of the maxima in the energy as a function of $J_2/J_1$ represents the
lower bound of the hatched region in Fig.~\ref{fig:pd} and Fig.~\ref{fig:pdz}.

 A further element of inconsistency of spin-wave theory is offered by the
 apparent violation of the Hellman-Feynman theorem
 \begin{equation}
 \langle S_x\rangle =  -\frac{\partial \langle {\cal H} \rangle}{\partial \Gamma}~.
 \label{e.HF}
 \end{equation}

\begin{figure}
\begin{center}
\rotatebox{270}{
\epsfig{file=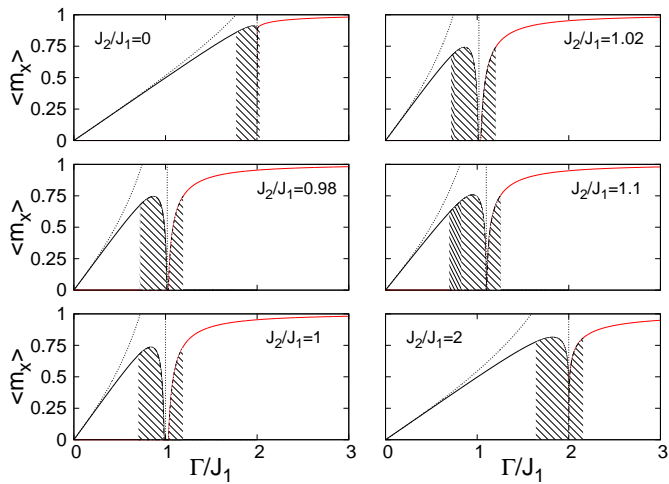,scale=0.32}
}
\null\vspace*{.5cm}
\caption{Field-induced magnetization as a function of the field for different values of $J_{2}/J_{1}$.
The highlighted regions are as in Fig.~\ref{fig:op}. The dotted lines correspond to the opposite of the derivatives of the average energies with respect to the magnetic field $\Gamma$. These curves deviate from the average induced
magnetization, signaling a violation of the Hellmann-Feynman theorem (see text). }
\label{fig:mx}
\end{center}
\end{figure}

\noindent The transverse magnetization, defined as $ \langle m_x \rangle = \langle S^x \rangle$ is
proportional to $m$ in the harmonic approximation : $ \langle S_x \rangle = S\sin\vartheta ~m$.
The deviation from Eq.~\eqref{e.HF} comes along with a strong non-monotonic
behavior, as shown in Fig.~\ref{fig:mx}.
This behavior is unphysical, implying a negative susceptibility.
This signals again that an ordered reference state does not lead to consistent results.
We identify therefore an additional region of strong anharmonic fluctuations with the
magnetization dip in the $m_{x}$ vs. $\Gamma$ curve. This dip lies inside the region
$\left[\Gamma_{1},\Gamma_{2}\right]$ where $\dfrac{\partial m_{x}}{\partial \Gamma}(\Gamma_{1})=0$
and $m_{x}(\Gamma_{2})=m_{x}(\Gamma_{1})$. The latter condition marks the upper
bound of the hatched region in Figs.~\ref{fig:pd} and \ref{fig:pdz}.

 Indeed Eq.~\eqref{e.HF} ceases to be strictly valid in the linear spin-wave approximation
 because in general the linear spin-wave Hamiltonian does \emph{not} have the form
 ${\cal H} = {\cal H}_0 - \Gamma S^{x}$. From inspection of Eqs.~\eqref{eq:lswham}
 and \eqref{reallsw} it is immediate to see that each term of the Hamiltonian
 depends on $\Gamma$ through $\theta(\Gamma)$, except for the polarized
 case in which $\theta = \pi/2$ independently of the value of $\Gamma$.
 In fact, to recover the identity of  Eq.~\eqref{e.HF} at order $1/S$, one has to include corrections
 to the harmonic ground state\cite{coletta2}.
 Even though Eq.~\eqref{e.HF} ceases to be an identity within linear spin-wave theory,
 the precision with which Eq.~\eqref{e.HF} is approximately verified can be used as
 a further criterion for the validity of the linear spin-wave approximation.
In Fig~\ref{fig:mx}, $\langle S_x \rangle$ is compared to $\partial \langle {\cal H} \rangle / \partial \Gamma$:
upon increasing the field, the quantities show a significant deviation from each other
when approaching the region already identified before as showing significant inconsistencies of spin-wave theory.

A strong violation of the Hellman-Feynman theorem Eq.~\eqref{e.HF}
is observed also in the case $J_2=0$.
In this limit the system reduces to the square-lattice Ising model
in a transverse field, which features a well-known quantum phase
transition (for a field $\Gamma_c/J_{1} \approx 1.5$  - see Ref.~\onlinecite{Blote}) between N\'eel order
and a quantum paramagnetic state. Interestingly the deviation between
$m_x$ and $\partial \langle {\cal H} \rangle / \partial \Gamma$ builds up
when approaching the quantum-critical field $\Gamma_c$, showing that
spin-wave theory is able to signal the quantum phase transition via
the breakdown of the consistency of its results with known theorems.


\section{Conclusions and discussion}
\label{conclusions}

 In this study we have applied a linear spin-wave theory analysis to the transverse-field
 Ising model on the checkerboard lattice, which represents a paradigmatic example of a
 frustrated Ising model with controlled quantum fluctuations. We find the remarkable
 result that harmonic quantum fluctuations are not able to lift the classical degeneracy,
 which is exponential in the linear dimension of the system when the next-to-nearest-neighbor
coupling $J_2$ exceeds the nearest-neighbor one $J_1$, and which is exponential in the system
size for $J_1 = J_2$, corresponding to the square-ice limit. This implies that spin wave theory
is inconclusive regarding the question of which classical ground state is selected by quantum effects, and that non-linear
quantum fluctuations play a central role in lifting the degeneracy in the exact ground state.
This result is consistent with existing studies of the square ice in a weak transverse field, for which
degenerate perturbation theory and numerics suggest a quantum-disordered ground state with the
structure of a resonating plaquette solid.

Our results suggest that a quantum-disordered ground
state persists beyond this limit.
Spin-wave theory indicates an anomalous inversion in the classical hierarchy
between reference states for strong fields and for $J_1 \sim J_2$, suggesting that
quantum fluctuations beyond the harmonic approximation can destroy classical order
in the system in that parameter range. Moreover spin-wave theory breaks down completely
close to the classical transition line to full polarization, where linear quantum fluctuations
are able to suppress all spin components. This identifies an interesting candidate region for
quantum-disordered ground states.

 Recent experimental developments have led to controlled realizations of frustrated Ising
 models, both in the classical limit and in the presence of tunable transverse fields. In particular
 the classical square-ice model, enriched with long-range dipolar interactions, is realized in recent
 experiments on square-lattice arrays of nano-patterned magnetic domains \cite{Wangetal06}.
 While quantum tunneling of the magnetization in these systems is not realistic,
 given the mesoscopic size of the magnetic moments, one can envisage scaling down the
 components of these artificial ice systems to single-molecule magnets \cite{Sessolietal06},
 which can be arranged into regular arrays on a surface \cite{Manninietal10}. 
 Transverse-field Ising models are currently realized by arrays of trapped ions \cite{Friedenaueretal08,
 Kimetal10, Islametal11}, where the internal states of the ions can be coupled with Ising
 Hamiltonians via virtual phonons, and where transverse fields are created by Raman
 laser schemes. The ions can be individually trapped by micro-traps, which can in turn be
 arranged into arbitrary planar arrays \cite{Schmiedetal09}, encompassing the checkerboard geometry
 explored here as well as other frustrated structures. Therefore the theoretical investigation of transverse-field
 Ising models on frustrated lattices is very compelling, as it promises to lead to the realization
 of novel quantum states in controlled artificial spin systems in the near future.

\begin{acknowledgments}
F. M. acknowledges the hospitality of the Ecole Normale Sup\'erieure de Lyon.
\end{acknowledgments}
\appendix

\section{Spin-wave theory for general Ising Hamiltonians in a transverse field
\label{diag}}

We present here several general formulas to study the quadratic quantum
fluctuation in a generic transverse field Ising system.
We consider a generic classical ground state with long-range magnetic
order, and with a magnetic unit cell containing $n$ spins. We denote
$S_{l,p}$ the {\it p}-th spin ($p=1\ldots n$) of the {\it l}-th cell. The most
general Hamiltonian supporting such a ground state has the form
\begin{equation}
	\mathcal{H}_{\rm TFI}=\dfrac{1}{2}\Sum_{lp,l'p'}{[J(\bm r_{l'}-\bm r_{l})]_{pp'}S_{l,p}^{z}S_{l',p'}^{z}}
	-\Gamma\Sum_{i}{S_{l,p}^{x}}~.
\end{equation}
Here ${\bm r}_l$ is the position of a reference site in the $l-$th unit cell, and
$J(\Delta {\bm r})$ is a $n\times n$ matrix containing the couplings between
spins in unit cells at a distance $\Delta{\bm r}$.

In the classical limit an applied transverse field rotates the $p$-th spins around
the $y$-axis by an angle $\vartheta_{p}$. As already discussed in Section
\ref{lswham},  we introduce a local rotation of the
spin configuration, $\tilde{{\bm S}}_{l,p} = \sigma_p \mathcal{R}_y (\sigma_{p}
\vartheta_{p}) {\bm S}_{l,p}$, where $\sigma_p = \pm 1$ is the orientation of the spin
in zero field. The rotation has the effect of reducing the $S=\infty$ ground state to a
perfectly ferromagnetic one. The classical energy of the $p$-th spin of each cell
has the expression
\[
\varepsilon_{cl,p}= \dfrac{S^{2}}{2}\sigma_{p}\cos\vartheta_{p}\Sum_{\Delta{\bm r},p'}
\left[J(\Delta{\bm r})\right]_{pp'}\sigma_{p'}\cos\vartheta_{p'}
-S\Gamma\sin\vartheta_{p},\\
\]
so that the total classical energy can be written as
\begin{equation}
	E_{cl} = \dfrac{N}{n}\Sum_{p}^{}{\varepsilon_{cl,p}}=\dfrac{N}{n}\Tr \varepsilon_{cl}
\end{equation}
where we have introduced the matrix $[\varepsilon_{cl}]_{p,p'} = \varepsilon_{cl,p} \delta_{p,p'}$.

We then consider small quantum fluctuations around this classical reference state, by
transforming the quantum spins to bosons via a linearised Holstein-Primakoff transformation
\cite{HP} valid in the limit of a small number of bosons $n_{l,p} \ll 2S$:
\begin{equation}
\tilde{S}^{z}_{l,p} = S-a_{l,p}^{\dagger}a_{l,p} ~~~~~ \tilde{S}^{x}_{l,p} \approx
\sqrt{\dfrac{S}{2}}\left(a_{l,p}^{\dagger}+a_{l,p}\right)
\end{equation}
Here $a_{l,p}$ and $a_{l,p}^{\dagger}$ are bosonic operators, satisfying
$[a_{l,p},a_{l,p}^{\dagger}]=1$ and $[a_{l,p}^{(\dagger)},a_{l,p}^{(\dagger)}]=0$.
The angles $\vartheta_{p}$ are chosen so that the classical reference state is stable. Thus
the linear terms in the bosonic operators vanish.

The quadratic Hamiltonian then reads
\begin{eqnarray}
		\mathcal{H}_{2}&=&E_{cl}+\Sum_{l,p}{\tilde{h}_{p}a^{\dagger}_{lp}a_{lp}} \\
		&&+\dfrac{1}{2}\Sum_{lp,l'p'}{\tilde{J}(\bm r_{l'}-\bm r_{l})_{pp'}\left(a^{\dagger}_{lp}+a_{lp}\right)
		\left(a^{\dagger}_{l'p'}+a_{l'p'}\right)}\nonumber
	\end{eqnarray}
 where
\begin{equation}
	\begin{array}{lcl}
	\tilde{h}_{p}&=&2\varepsilon_{cl,p}/S+\Gamma\sin\vartheta_{p}\\
	\tilde{J}(\Delta\bm r)_{pp'}&=&J(\Delta\bm r)_{pp'}\sin\vartheta_{p}\sin\vartheta_{p'}
	\end{array}
\end{equation}
We then introduce the Fourier transform of the bosonic operators and of the interaction
\begin{equation}
	\begin{array}{lcl}
		a_{\kb,p}&=&\sqrt{\dfrac{2}{N}}\Sum_{l}{\e^{\ii \kb.{\bm r}_l}a_{l,p}}\\
		J(\kb)&=&\Sum_{l}{\e^{-\ii \kb.\Delta\bm r}\tilde{J}(\Delta\bm r)}~.
	\end{array}
\end{equation}
The quadratic Hamitonian can then be written in the compact form
\begin{equation}
	\mathcal{H}_{2} = \dfrac{N}{n}\Sum_{p}{\left(\varepsilon_{cl,p}-\dfrac{\tilde{h}_{p}}{2}\right)}
	+\dfrac{1}{2}\Sum_{\kb}{A_{\kb}^{\dagger}M_{\kb}A_{\kb}}
\end{equation}
where
\begin{equation}
	\begin{array}{lcl}
		\tilde{h}_{p}&=&\dfrac{2}{S}~\varepsilon_{cl,p}+\Gamma\sin\vartheta_{p}\\
		A_{\kb}^{\dagger} &=& ( a_{\kb,1}^{\dagger},\ldots,a_{\kb,n}^{\dagger},a_{-\kb,1},\ldots,a_{-\kb,n} )\\
		M_{\kb}&=&	\begin{pmatrix} 	\Delta_{\kb}&\Delta_{\kb}\\
														\Delta_{\kb}&\Delta_{\kb}
							\end{pmatrix}
							-\begin{pmatrix}
								\varepsilon_{cl}&0_{n}\\
								0_{n}&\varepsilon_{cl}
							\end{pmatrix}\\
		\Delta_{\kb}&=&\dfrac{1}{2}\left(\tilde{J}(\kb) +\tilde{J}(\kb)^{\dagger}\right)
	\end{array}
\end{equation}
This Hamiltonian can be diagonalized by a $n$-mode Bogolyubov transformation.
This consists in finding the transformation $A_{\kb}=T_{\kb}B_{\kb}$,
with $B_{\kb}=( b_{\kb,1}^{\dagger},\ldots,b_{\kb,n}^{\dagger} ,b_{-\kb,1},\ldots,
b_{-\kb,n} )^{T}$, such that $A_{\kb}^{\dagger} M_{\kb}A_{\kb} = \Sum_{p}
{\omega_{\kb}^{(p)}b_{\kb,p}^{\dagger}b_{\kb,p}}$ and $[b_{\kb,p},
b_{\kb,p}^{\dagger}]=1$ and $[b_{\kb,p}^{(\dagger)},b_{\kb,p}^{(\dagger)}]=0$.

We introduce the matrix $\Sigma$, given by
\[
\Sigma=\begin{pmatrix}	I_{n}	&	0_{n}\\	0_{n}	&	-I_{n}	\end{pmatrix} ~,
\]
\noindent the matrix $Z_{\kb}$ of the right eigenvectors of $\Sigma M_{\kb}$,
and the unitary matrix $U_{\kb}$ such that $U_{\kb}^{\dagger}Z_{\kb}^{\dagger}
\Sigma Z_{\kb}U_{\kb}=\diag(l_{\kb}^{(1)},\ldots,l_{\kb}^{(n)})=L_{\kb}$.
The transformation matrix $T_{\kb}$ is then obtained as \cite{WesselM04, Muccioloetal04, BlaizotR85}
\begin{equation}
	T_{\kb}=Z_{\kb}U_{\kb}|L_{\kb}|^{\mbox{-}\nicefrac{1}{2}}~.
\end{equation}
In particular, the eigenmodes $\omega_{\kb}^{(p)}$ are the eigenvalues of $\Sigma M_{\kb}$.

If the matrices $\Delta_{\kb}$ and $\varepsilon_{cl}$ commute (which is the case for the N\'eel and
collinear states of the checkerboard Ising model studied here, having $\varepsilon_{cl}=\varepsilon_{0}I_{n}$), the eigenmodes
$\omega_{\kb}^{(p)}$ can be expressed in terms of the eigenvalues $\lambda_{\kb}^{(p)}$ of $\Delta_{\kb}$
in the form
\begin{equation}
	\omega_{\kb}^{(p)}=\dfrac{\varepsilon_{cl,p}}{2}\sqrt{1+4\dfrac{\lambda_{\kb}^{(p)}}{\varepsilon_{cl,p}}}~.
\end{equation}


\section{Energy and magnetization in the case of N\'eel and collinear states
\label{diagNc}}

As mentioned above,
in the particular case of the N\'eel and collinear states of the checkerboard lattice Ising model
studied in this work,
the classical energies $\varepsilon_{cl,p}$ are all equal to $\varepsilon_{0}$. The
mean energy of the system then reads
\begin{equation}
\begin{array}{lcl}
 \langle E \rangle &=& N(\varepsilon_{cl} -\varepsilon_0/2) +\dfrac{1}{2}\Sum_{\kb,p}{\omega_{\kb,p}}\\
 &=&N\varepsilon_{cl}+\dfrac{\varepsilon_{0}}{4}\Sum_{\kb,p}{\left(\sqrt{1+4\dfrac{\lambda_{\kb,p}}{\varepsilon_{0}}}-1\right)}
\end{array}
\end{equation}
As $\Delta_{\kb}$ is proportional to $\Gamma^{2}$, so are its eigenvalues.
We can then expand the mean energy per spin in powers of $\Gamma^2$.
We will introduce rescaled eigenvalues $\tilde{\lambda}_{\kb,p}$ defined as
 $\lambda_{\kb,p}=\varepsilon_{0}\Gamma^{2}\tilde{\lambda}_{\kb,p}$.
\begin{equation}
	\begin{array}{lcl}
	\langle\varepsilon\rangle&=&\varepsilon_{cl}
	+\dfrac{\varepsilon_{0}}{N}\Sum_{\kb}\Sum_{m=1}^{\infty}{\alpha_{m}\tilde{\lambda}_{\kb,p}^{m}\Gamma^{2m}}\\
	 &=&\varepsilon_{cl}+\dfrac{\varepsilon_{0}}{N}\Sum_{\kb,m}{\alpha_{m}\Tr\left(\dfrac{\Delta_{\kb}}{\varepsilon_{0}}\right)^{m}}
	\end{array}
	\label{eq:expansion}
\end{equation}
where the $\alpha_{m}$ are defined by $	\sqrt{1+4x}-1=
4\Sum_{m=1}^{\infty}{\alpha_{m}x^{m}}$.
In the case of a N\'eel or a collinear case, the trace of $\Delta_{\kb}$
averages to zero in the Brillouin zone. Thus the first non-zero correction
to the classical energy in Eq.~\eqref{eq:expansion} is
of fourth order in $\Gamma$.

In both cases, if we introduce the ratio $J=J_{2}/J_{1}$, we get
\begin{equation}
\langle \varepsilon \rangle = \varepsilon_{cl}+\Sum_{p>1}{c_{p}(J)
		\left(\dfrac{J_{1}}{S(\nu_{2}-J)^{2}}\right)^{p}\Gamma^{2p}}
\end{equation}
where the $c_{p}$ coefficients only depend on $J$, and not
on the considered state. It is then obvious that the expansion becomes
independent of the classical state if $J=1$. Thus the classical degeneracy
of the ice model is not lifted by harmonic fluctuations.

Similarly, if all the $\omega_{\kb,p}$ are real (which is the case whenever spin-wave
theory holds), we have

\begin{equation}
\begin{array}{lcl}
 m &=& 1 - \dfrac{1}{NS}\Sum_{\kb,p}{\left\langle a^{\dagger(p)}_{\kb}a_{\kb}^{(p)}\right\rangle}\\
 &=& 1 - \dfrac{1}{4NS}\Sum_{\kb,p}{\left(\dfrac{2\omega_{\kb,p}}{\varepsilon_{0}}
 +\dfrac{\varepsilon_{0}}{2\omega_{\kb,p}}-2\right)}\\
 &=& 1 - \dfrac{1}{NS}\Sum_{m=1}^{\infty}\Sum_{\kb,p}{\beta_{m}\widetilde{\lambda_{\kb,p}}^{m}\Gamma^{2m}}
\end{array}
\end{equation}
where the $\beta_{m}$ are defined by $	\sqrt{1+4x}+\dfrac{1}{\sqrt{1+4x}}-2=
4\Sum_{m=1}^{\infty}{\beta_{m}x^{m}}$.



\begin{thebibliography}{99}
%


\bibitem{Milaetal2011} \emph{Introduction to Frustrated Magnetism: materials, experiments, theory},
C. Lacroix, F. Mila and P. Mendels (Eds.), Springer, Berlin, 2011.
\bibitem{Balents10} L. Balents, Nature, {\bf 464}, 199 (2010)
\bibitem{Whiteetal11} Nonetheless a spin-liquid ground state has been recently reported for the
kagom\'e lattice Heisenberg antiferromagnet, see S. Yan,  D. A. Huse, and S. R. White,
Science {\bf 332}, 1173 (2011).
\bibitem{Hermeleetal04} M. Hermele, M. P. A. Fisher, and L. Balents, Phys. Rev. B {\bf 69}, 064404 (2004).
\bibitem{Isakovetal06} S. V. Isakov, Y. B. Kim, and A. Paramekanti,
Phys. Rev. Lett. {\bf 97}, 207204 (2006).
\bibitem{Banerjeeetal08} A. Banerjee, S. V. Isakov, K. Damle, and Y. B. Kim, Phys. Rev. Lett.
{\bf 100}, 047208 (2008).
\bibitem{Isakovetal11} S. V. Isakov, M. B. Hastings, and R. G. Melko,
Nature Physics, doi:10.1038/nphys2036 (2011).
\bibitem{MoessnerQDM} R. Moessner and K. S. Raman in \emph{Introduction to Frustrated Magnetism: materials, experiments, theory},
C. Lacroix, F. Mila and P. Mendels (Eds.), Springer, Berlin, 2011.
\bibitem{Sikoraetal09} O. Sikora, F. Pollmann, N. Shannon, K. Penc and P. Fulde, Phys. Rev. Lett. {\bf 103}, 247001 (2009).
\bibitem{MoessnerS01-2} R. Moessner and S. L. Sondhi, Phys. Rev. Lett. {\bf 86}, 1881 (2001).
\bibitem{MoessnerS01} R. Moessner and S. L. Sondhi, Phys. Rev. B {\bf 63}, 224401 (2001).
\bibitem{Chakrabartietal96} B. K. Chakrabarti, A. Dutta, P. Sen, \emph{Quantum Ising Phases and Transitions
in Transverse Field Ising Models}, Springer, Berlin, 1996.
\bibitem{Lieb67} E. H. Lieb, Phys. Rev. {\bf 162}, 162 (1967).
\bibitem{CastroNetoetal06} A. H. Castro Neto and P. Pujol and E. Fradkin, Phys. Rev. B {\bf 74}, 024302 (2006);
\bibitem{Moessneretal04} R. Moessner, O. Tchernyshyov, and S. Sondhi, J. Stat. Phys. {\bf 116}, 755 (2004).
\bibitem{Shannonetal04} N. Shannon, G. Misguich, and K. Penc, Phys. Rev. B {\bf 69}, 220403(R) (2004).
\bibitem{Zhou05} Y. Zhou, Phys. Rev. B {\bf 72}, 205116 (2005).
\bibitem{Ralkoetal10} A. Ralko, F. Trousselet, and D. Poilblanc, Phys. Rev. Lett. {\bf 104}, 127203 (2010).
\bibitem{harris}A.~B. Harris, C. Kallin and A.~J. Berlinsky, Phys. Rev. B {\bf 45}, 2899 (1992).
\bibitem{chalker} J.~T. Chalker, P.~C.~W. Holdsworth, and E.~F. Shender, Phys. Rev. Lett. {\bf 68}, 855 (1992).
\bibitem{ritchey} I. Ritchey, P. Chandra, and P. Coleman, Phys. Rev. B {\bf 47}, 15342 (1993).
\bibitem{KorshunovDoucotPhysRevLett.93.097003} S.~E. Korshunov and B. Dou\c cot, Phys. Rev. Lett. {\bf 93}, 097003 (2004).
\bibitem{Korshunov05} S.~E. Korshunov, Phys. Rev. B {\bf 71}, 174501 (2005); Phys. Rev. Lett. {\bf 94}, 087001 (2005).
\bibitem{henley} C.~L. Henley, Phys. Rev. Lett. {\bf 96},  047201  (2006).
\bibitem{coletta} T. Coletta, J.-D. Picon, S.E. Korshunov, F. Mila, Phys. Rev. B 83, 054402 (2011).
\bibitem{MuellerHartmannR02} E. M\"uller-Hartmann and A. Reischl Eur. Phys. J. B {\bf 28}, 173 (2002).
\bibitem{misguich} G. Misguich and F. Mila, Phys. Rev. B 77, 134421 (2008).
\bibitem{HP} T. Holstein and H. Primakoff, Phys. Rev. {\bf 58}, 1098 (1940)
\bibitem{Henryprep} L.-P. Henry and T. Roscilde, in preparation.
\bibitem{CMS} C. Castelnovo, and R. Moessner, and S. L. Sondhi, Nature {\bf 451}, 42 (2008).
\bibitem{coletta2} T. Coletta, N. Laflorencie, F. Mila, unpublished (arXiv:1112.5586).
\bibitem{Blote} H. W. J. Bl\"ote and Y. Deng, Phys. Rev. E {\bf 66}, 066110 (2002).
\bibitem{Wangetal06} R. F. Wang, C. Nisoli, R. S. Freitas, J. Li, W. McConville, B. J. Cooley, M. S. Lund, N. Samarth, C. Leighton, V. H. Crespi, and P. Schiffer, Nature {\bf 439}, 303 (2006).
\bibitem{Sessolietal06} D. Gatteschi, R. Sessoli, and J. Villain, \emph{Molecular Nanomagnets}, Oxford University Press, 2006.
\bibitem{Manninietal10} M. Mannini, F. Pineider, P. Sainctavit, C. Danieli, E. Otero, C. Sciancalepore, A. M. Talarico, M.-A. Arrio, A. Cornia, D. Gatteschi, and R. Sessoli, Nature Materials {\bf 8}, 194 (2009).
\bibitem{Friedenaueretal08} A. Friedenauer, H. Schmitz, J. Gl\"uckert, D. Porras and T. Sch\"atz, Nature Physics {\bf 4}, 757 (2008).
\bibitem{Kimetal10} K. Kim, M.-S. Chang, S. Korenblit, R. Islam, E. E. Edwards, J. K. Freericks, G.-D. Lin, L.-M. Duan, and C. Monroe, Nature {\bf 465}, 590 (2010).
\bibitem{Islametal11} R. Islam, E. E. Edwards, K. Kim, S. Korenblit, C. Noh, H. Carmichael, G.-D.Lin, L.-M. Duan, C.-C. Joseph Wang, J. K. Freericks, and C. Monroe, Nature Communications {\bf 2}, 377 (2011).
\bibitem{Schmiedetal09} R.Schmied, J. H. Wesenberg and D. Leibfried, Phys. Rev. Lett. {\bf 102}, 233002 (2009).
\bibitem{Muccioloetal04} E. R. Mucciolo, A. H. Castro Neto, and C. Chamon, Phys. Rev. B {\bf 69}, 214424 (2004).
\bibitem{WesselM04} S. Wessel and I. Milat, Phys. Rev. B {\bf 71}, 104427 (2005).
\bibitem{BlaizotR85} J.-P. Blaizot and G. Ripka, \emph{Quantum Theory of Finite Systems}, MIT Press, 1985.

\end{thebibliography}
\end{document}